\documentclass[a4paper, twoside, 11pt]{llncs}
\usepackage{amsmath}
\usepackage{fullpage}
\usepackage{algorithm}
\usepackage{algorithmic}

\newcommand{\ACo}{\mbox{{\sf AC}$^1$}}
\newcommand{\NCo}{\mbox{{\sf NC}$^1$}}
\newcommand{\Log}{\mbox{{\sf L }}}
\newcommand{\ULcoUL}{\mbox{{\sf UL $\cap$ coUL }}}

\newcommand{\NL}{\mbox{{\sf NL }}}
\newcommand{\PL}{\mbox{{\sf PL }}}
\newcommand{\NP}{\mbox{{\sf NP }}}
\usepackage{amssymb}
\usepackage{latexsym}

\title{$3$-connected Planar Graph Isomorphism is in Log-space}

\author{Samir Datta \inst{1}\and Nutan Limaye \inst{2} \and Prajakta
Nimbhorkar \inst{2}}
\authorrunning{S. Datta \and N. Limaye \and P. Nimbhorkar}

\institute{ Chennai Mathematical Institute, Chennai 603 103, India.\\
   \email{sdatta@cmi.ac.in}
\and	The Institute of Mathematical Sciences, Chennai 600 113,
  India. \\ \email{{nutan,prajakta}@imsc.res.in}}

\begin{document}
\bibliographystyle{plain}
\maketitle
\begin{abstract}
We show that the isomorphism of $3$-connected planar graphs can be 
decided in deterministic log-space. This improves the previously known bound
\ULcoUL\ of \cite{TW07}. 
\end{abstract}

\section{Introduction}
The general graph isomorphism problem is a well studied problem in
computer science. Given two graphs, it deals with finding a bijection
between the sets of vertices of these two graphs, such that the
adjacencies are preserved. The problem is in \NP, but it is not known to
be complete for \NP. In fact, it is known that if it is complete for \NP,
then the polynomial hierarchy collapses to its second level. On the
other hand, no polynomial time algorithm is known. For general graph
isomorphism \NL\ and \PL\ hardness is known \cite{Toran00}, whereas for
trees, \Log\ and \NCo\ hardness is known, depending on the encoding of the
input \cite{MJT98}.

\par In literature, many special cases of this general graph isomorphism
problem have been studied. In some cases like trees \cite{Lin92},
\cite{Buss97}, or graphs with coloured vertices and bounded colour
classes \cite{Luks86}, NC algorithms are known. We are interested in the 
case where the
graphs under consideration are planar graphs. In \cite{Wei66}, Weinberg
presented an $O(n^2)$ algorithm for testing isomorphism of $3$-connected
planar graphs. Hopcroft and Tarjan \cite{HT74} extended
this for general planar graphs, improving the time complexity to
$O(n\log n)$. Hopcroft and Wong \cite{HW74} further improved it to give
a linear time algorithm. Its parallel complexity was first considered by
Miller and Reif \cite{MR91} and Ramachandran and Reif \cite{RR94}. 
They gave an upper bound of \ACo. Recently Thierauf and Wagner \cite{TW07}
improved it to \ULcoUL\ for $3$-connected planar graphs. They also proved
that this problem is hard for \Log. In this paper, we give a log-space
algorithm for $3$-connected planar graph isomorphism, thereby settling
its complexity.

\par Thierauf and Wagner use shortest paths between nodes of a graph to
obtain a canonical spanning tree. A systematic traversal of this tree
generates a canonical form for the graph. The best known upper bound for
shortest paths in planar graphs is \ULcoUL\ \cite{TW07}. Thus the total
complexity of their algorithm goes to \ULcoUL, despite the fact that all
other steps can be done in \Log. 
\par We identify that their algorithm hinges on making a systematic
traversal of the graph in canonical way. Thus we bypass the step of
finding shortest paths and give an orthogonal approach for finding such
a traversal. We use the notion of universal exploration sequences (UXS) 
defined in \cite{koucky01}. Given a graph on $n$ vertices with maximum 
degree $d$, a UXS is nothing but a polynomial length string over 
$\{0,\ldots ,d-1\}$. Such a sequence can be used to traverse the graph
for a chosen combinatorial embedding $\rho$, starting
vertex $u$ and a starting edge $e=\{u,v\}$. 
Reingold \cite{Rei05} proved that such a universal
sequence can be constructed in \Log. Using this result, we canonize a
$3$-connected planar graph in log-space. To our knowledge, this is the
best upper bound for this class of graphs. 
\par In Section \ref{sec:prelim}, we give some basic definitions of
complexity classes and formally define the notion of universal exploration
sequences. In Section \ref{sec:alg}, we describe our log-space
algorithm. We conclude with a discussion of open problems in Section 
\ref{sec:concl}.

\section{Preliminaries}\label{sec:prelim}
In this section, we give a brief introduction of the graph isomorphism
problem and the notion of universal exploration sequences.

\subsection{Universal Exploration Sequences}
Let $G=(V,E)$ be a $d$-regular graph, with given combinatorial embedding
$\rho$. The edges around any vertex $u$ can be numbered $\{0,1,\ldots ,
d-1\}$ according to $\rho$ arbitrarily in clockwise order. 
A sequence $\tau_1 \tau_2
\ldots \tau_k \in \{0,1,\ldots ,d-1\}^k$ and a starting edge
$e_0=(v_{-1},v_0)\in E$, define a walk $v_{-1},v_0,\ldots v_k$ as
follows: For $0\leq i \leq k$, if $(v_{i-1},v_i)$ is the $s^{th}$
edge of $v_i$, let $e_i=(v_i,v_{i+1})$ be $(s+\tau_i)^{th}$ edge of $v_i$
modulo $d$. 

\begin{definition}
Universal Exploration sequences (UXS): A sequence $\tau_1\tau_2\ldots
\tau_l\in \{0,1,\ldots d-1\}^l$ is a universal exploration sequence for
$d$-regular graphs of size at most $n$ if for every connected
$d$-regular graph on at most $n$ vertices, any numbering of its edges,
and any starting edge, the walk obtained visits all the vertices of the
graph.
\end{definition}
Following lemma suggests that UXS can be constructed in \Log\cite{Rei05}:
\begin{lemma}
There exists a log-space algorithm that takes as input $(1^n,1^D)$
and produces an $(n,D)$-universal exploration sequence.
\end{lemma}

\subsection{The Graph Isomorphism Problem}

\begin{definition}
Graph isomorphism: Two graphs $G_1=(V_1,E_1)$ and $G_2=(V_2,E_2)$ are 
said to be isomorphic if there is a bijection $\phi:V_1\rightarrow V_2$
such that $(u,v)\in E_1$ if and only if $(\phi(u),\phi(v))\in E_2$.
\end{definition}

Let GI be the problem of finding such a bijection
$\phi$ given two graphs $G_1,G_2$. Let Planar-GI be the special case of
GI when the given graphs are planar.
$3$-connected planar graph isomorphism
problem is a special case of Planar-GI when the graphs are $3$-connected
planar graphs. We recall the definition of $3$-connected planar graphs
here:
\par A graph $G$ is connected if there is a path between any two vertices in
G. A vertex $v\in V$ is an articulation point if $G-v$ is not connected. A
pair of vertices $u,v \in V$ is a separation pair if
$G(V\setminus\{u,v\})$ is not connected.
A biconnected graph contains no articulation points. A $3$-connected graph
contains no separation pairs.

\section{Log-space Algorithm for $3$-connected Planar-GI}\label{sec:alg}
In this section, we prove following theorem:
\begin{theorem}\label{thm:isoL}
Given two $3$-connected planar graphs $G$ and $H$, deciding whether $G$
is isomorphic to $H$ can be done in \Log.
\end{theorem}

\par For general planar graphs, the best known parallel algorithm
runs in \ACo\ \cite{MR91}. Thierauf and Wagner \cite{TW07} recently
improved the bound for the case of $3$-connected planar graphs to
\ULcoUL. This case is easier due to a result by Whitney \cite{Whi33}
that every planar
3-connected graph has precisely two embeddings on a sphere, where one
embedding is the mirror image of the other. Moreover, one can efficiently
compute these embeddings. 
\par Using these embeddings, Thierauf and Wagner 
compute a code for a graph, such that isomorphic graphs will have the same
code. A code with this property is called a canonical code for the graph.
They construct it via a spanning
tree, which depends upon the planar embedding of the graph. Bourke, Tewari
and Vinodchandran \cite{BTV07} proved that planar reachability is in
\ULcoUL. Thierauf and Wagner extend their result for computing distances 
in planar graphs in \ULcoUL\ and crucially use this in the
construction of the spanning tree. Once this spanning tree is
constructed, a canonical code can be obtained in \Log.
\par Our approach bypasses the spanning tree construction step
and thus eliminates distance computations. In that sense, we believe 
that this is a completely new approach for computing canonical codes for
$3$-connected planar graphs.
\par Our algorithm can be outlined as follows: 
\begin{enumerate}
\item Given a $3$-connected planar graph $G=(V,E)$, find its planar
embedding $\rho$. 
\item Make the graph $3$-regular canonically for this embedding $\rho$
to obtain an edge-coloured graph $G'$ as described in algorithm
\ref{alg:3reg}.
\item Find the canon of $G'$ using algorithm \ref{alg:canon}.
\end{enumerate}
The step $1$ is in log-space due to a result by Allender and Mahajan
\cite{AM00}. We prove that steps $2$ and $3$ can also be done in
log-space. Step $3$ uses the idea of UXS introduced by Koucky \cite{}.
Step $2$ essentially does the preprocessing in order to make step $3$
applicable. 

\par The canonical code thus constructed is specific to the choice
of the combinatorial embedding, the starting edge, and the starting vertex.
Given two graphs $G$ and $H$, we fix these arbitrarily for $G$ and cycle
through both embeddings and all choices of the starting edge and the 
starting vertex for $H$, comparing the codes for each of them. As there are 
only polynomially many choices, this loop runs in \Log.

\subsection{Making the graph $3$-regular}
In this section, we describe the procedure to make the graph
$3$-regular. In Section \ref{sec:canon}, we use Reingold's construction
for UXS \cite{Rei05} to come up with a canonical code. As Reingold's 
construction \cite{Rei05} for UXS requires
the graph to have constant degree, we do this preprocessing step. In Lemma
\ref{lemma:3reg}, we prove that two given graphs are isomorphic if and only 
if they are isomorphic after the preprocessing step. We note that after the
preprocessing step, the graph does not remain $3$-connected, however,
the embedding of the new graph is inherited from the given graph. Hence
even the new graph has only two possible embeddings.
\par We now describe the preprocessing steps in Algorithm
\ref{alg:3reg}. Note that the new graph thus obtained has $2|E|$
vertices.

\begin{algorithm}\label{alg:3reg}
\caption{Procedure to get a $3$-regular planar graph $G'$ from
$3$-connected planar graph $G$.}
\begin{algorithmic}
\STATE Input: A $3$-connected planar graph $G$ with planar combinatorial
embedding $\rho$.
\STATE Output: A $3$-regular planar graph $G'$ on $2m$ vertices, with 
edges coloured $1$ and $2$ and planar combinatorial embedding $\rho'$. 
\FORALL{$v_i\in V$}
\STATE Replace $v_i$ of by a cycle $\{v_{i1},\ldots ,v_{id_i}\}$ on $d_i$ 
vertices, where $d_i$ is the degree of $v_i$. 
\STATE The $d_i$ edges 
$\{e_{i1},\ldots ,e_{id_i}\}$ incident to $v_i$ in $G$ are now incident to 
$\{v_{i1},\ldots ,v_{id_i}\}$ respectively.
\STATE Colour the cycle edges with colour $1$.
\STATE Colour $e_{i1},\ldots ,e_{id_i}$ by colour $2$.
\ENDFOR
\end{algorithmic}
\end{algorithm}

\begin{lemma}\label{lemma:3reg}
Given two $3$-connected planar graphs $G_1,G_2$, $G_1\cong G_2$ if and
only if $G_1'\cong G_2'$ where the isomorphism between $G_1'$ and $G_2'$
respects colours of the edges.
\end{lemma}

\begin{proof} Let $G_1=(V_1,E_1)$ and $G_2=(V_2,E_2)$ be two $3$-connected planar
graphs with planar combinatorial embeddings $\rho_1$ and $\rho_2$
respectively. Let $\phi:V_1 \rightarrow V_2$ be an isomorphism between 
the oriented graphs $(G_1,\rho_1)$ and $(G_2,\rho_2)$. By isomorphism of
oriented graphs we mean that the graphs are isomorphic for the fixed
embeddings, in our case $\rho_1$ and $\rho_2$. Construct $G_1'$
and $G_2'$ preserving the orientation of original edges from $G_1$ and
$G_2$ respectively. Let the orientations be $\rho_1'$ and $\rho_2'$.
By our construction, edges around a vertex in $G_1 
(\textrm{respectively }G_2)$ get the same 
combinatorial embedding around the corresponding cycle in $G_1'$ $(G_2')$.
Consider an edge $\{v_i,v_j\}$ in $E_1$. Let $\phi(v_i)=u_k$ and
$\phi(v_j)=u_l$. $\{u_k,u_l\}\in E_2$. Let corresponding edge in $G_1'$
be $\{v_{i_p},v_{i_q}\}$ and that in $G_2'$ be $\{u_{k_r},u_{k_s}\}$.
Then define a map $\phi':V_1'\rightarrow V_2'$ such that
$\phi'(v_{i_p})=u_{k_r}$ and $\phi'(v_{j_q})=u_{k_s}$. It is easy to see
that $\phi'$ is an isomorphism for edge-coloured oriented graphs 
$(G_1',\rho_1')$ and $(G_2',\rho_2')$.
\par Now let $\phi'$ be an isomorphism between oriented graphs 
$(G_1',\rho_1')$ and
$(G_2',\rho_2')$. Let $e=\{v_{i_p},v_{i_q}\}\in E_1'$ where $v_{i_p}$ and
$v_{i_q}$ correspond to the same vertex $v_i$ in $G_1$. Then
$colour(e)=1$ and $e'=\{\phi'(v_{i_p}),\phi'(v_{i_q})\}\in E_2'$ and
$colour(e')=1$. Thus $\phi'$ maps copies of the same vertex of $G_1$ to
copies of a single vertex of $G_2$. Hence a map $\phi$ can be derived
from $\phi'$ in a natural way. It is easy to see that $\phi$ is an
isomorphism between oriented graphs $(G_1,\rho_1)$ and $(G_2,\rho_2)$.
\end{proof}

\subsection{Obtaining the canonical code}\label{sec:canon}
\par Lemma~\ref{lemma:3reg} suggests that for given embeddings $\rho_1$,
$\rho_2$ of $G_1$
and $G_2$, it suffices to check the $3$-regular oriented graphs
$(G_1',\rho_1')$ and $(G_2',\rho_2')$
for isomorphism. This can be done as follows:

\begin{algorithm}\label{alg:canon}
\begin{algorithmic}
\STATE Input: Edge-coloured graph $G=(V,E)$ with maximum degree $3$ and
combinatorial embedding $\rho$, starting vertex $v$, starting edge
$e=(u,v)$
\STATE Output: canon of $G$.
\STATE Construct a $(n,3)$ universal exploration sequence $U$.
\STATE With starting vertex $v\in V$ and edge $e=(u,v)$ incident
to it, traverse $G$ according to $U$ and $\rho$ and output the labels of the
vertices. 
\STATE Give labels to the vertices according to their first occurrence in 
this output sequence.
\STATE For every $(i,j)$ in this labelling, output whether $(i,j)$ is an
edge or not. If it is an edge, output its colour. This gives a canon for 
the graph.
\end{algorithmic}
\caption{Procedure $canon(G,\rho, v,e=(u,v))$}
\end{algorithm}

\begin{lemma}
\label{lemma:canon}
Let $\sigma_1=canon(G_1',\rho_1',v_1,e_1=(u_1,v_1))$ and
$\sigma_2=canon(G_2',\rho_2',v_2,e_2=(u_2,v_2))$. If $\sigma_1=\sigma_2$
then $G_1'\cong G_2'$. Further, if $G_1'\cong G_2'$ then for some choice
of $\rho_2',v_2,e_2$, $\sigma_1=\sigma_2$.
\end{lemma}
\begin{proof}
If $G_1'\cong G_2'$, then there is
a bijection $\phi:V_1'\rightarrow V_2'$ for corresponding embeddings
$\rho_1',\rho_2'$. Let $e_1=(u,v)\in E_1'$. Then
$e_2=(\phi(u),\phi(v))\in E_2'$. Let $e_1$ and $e_2$ be chosen as
starting edges and $v$ and $\phi(v)$ as starting vertices
for traversal using UXS $U$ for  $(G_1',\rho_1')$ and $(G_2', \rho_2')$
respectively. Let $T_1$ and $T_2$ be the output sequence. If a vertex 
$w\in V_1'$
occurs at position $l$ in $T_1$ then $\phi(w)\in V_2'$ occurs at
position $l$ in $T_2$. Thus the sequences are canonical when projected
down to the first occurrences and hence $\sigma_1=\sigma_2$.
\par Let $\sigma_1=\sigma_2=\sigma$. The labels of vertices in $\sigma$
are just a relabelling of vertices of $V_1'$ and $V_2'$. These
relabellings are some permutations, say $\pi_1$ and $\pi_2$. 
Then $\pi_1\cdot
\pi_2^{-1}:V_1'\rightarrow V_2'$ is a bijection.
\end{proof}

\par After constructing canonical code $\sigma'$ for a graph $G'$, 
it remains to construct canonical code $\sigma$ for the original graph $G$.
For each edge $(i,j)$ of colour $2$ in $\sigma'$, traverse along the
edges coloured $1$ starting from $i$ and find the minimum among the 
vertices visited. Let it be $p$. Repeat the process for $j$. Let the minimum
vertex visited along edges of colour $1$ be $q$. Output the edge
$(p,q)$. The sequence thus obtained contains $n$ labels for vertices,
each between $\{1,2,\ldots ,2m\}$. This can further be converted into a
sequence with labels for vertices between $\{1,2,\ldots ,n\}$ by finding
the rank of each of the labels. This gives us $\sigma$. Correctness follows from the fact that
vertices connected with edges of colour $1$ are copies of the same
vertex in $G$, hence they should get the same number. 
\par 
Clearly, each of the above steps can be performed in \Log and hence
the algorithm runs in \Log.
This proves Theorem \ref{thm:isoL}.

\section{Conclusion}\label{sec:concl}
Our note settles the open question mentioned in \cite{TW07} by giving a
log-space algorithm for $3$-connected planar graph isomorphism. The
most challenging question is to settle the complexity of the general
graph isomorphism problem. The other important goal is to improve upon
the \ACo\ upper bound of \cite{MR91} for planar graph isomorphism. 

\section{Acknowledgment}
We thank Jacobo Toran, V. Arvind, and Meena Mahajan for helpful
discussions.
\bibliography{iso}
\end{document}